\documentclass[sigconf]{acmart}
\AtBeginDocument{%
  }

\setcopyright{cc}
\setcctype{by}
\copyrightyear{2026}
\acmYear{2026}
\acmConference[KDD '26]{Proceedings of the 32nd ACM SIGKDD Conference
  on Knowledge Discovery and Data Mining V.2}{August 09--13,
  2026}{Jeju Island, Republic of Korea}
\acmBooktitle{Proceedings of the 32nd ACM SIGKDD Conference on Knowledge
  Discovery and Data Mining V.2 (KDD '26), August 09--13, 2026, Jeju
  Island, Republic of Korea}
\acmISBN{979-8-4007-2259-2/2026/08}
\acmDOI{10.1145/3770855.3818340}
\newcommand{\CLIMBER}{Climber-Pilot}

\usepackage{booktabs} % 提供了 \toprule, \midrule, \bottomrule 美化表格线
\usepackage{multirow} % 提供了 \multirow 用于合并行
\usepackage{enumitem}
\usepackage{balance}  % 提供 \balance 平衡末页两列
\usepackage{bm}       % 提供 \bm 用于数学公式中的加粗

\begin{document}

%%
%% The "title" command has an optional parameter,
%% allowing the author to define a "short title" to be used in page headers.
\title{\text{\CLIMBER}: A Non-Myopic Generative Recommendation Model Towards Better Instruction-Following}

%%
%% The "author" command and its associated commands are used to define
%% the authors and their affiliations.
%% Of note is the shared affiliation of the first two authors, and the
%% "authornote" and "authornotemark" commands
%% used to denote shared contribution to the research.
% \author{Da Guo}
% % \authornote{Both authors contributed equally to this research.}
% \email{trovato@corporation.com}
% \orcid{1234-5678-9012}
% \author{G.K.M. Tobin}
% % \authornotemark[1]
% \email{webmaster@marysville-ohio.com}
% \affiliation{%
%   \institution{Institute for Clarity in Documentation}
%   \city{Dublin}
%   \state{Ohio}
%   \country{USA}
% }

\author{Da Guo}
\affiliation{%
  \institution{NetEase Cloud Music}
  \city{Hangzhou}
  \country{China}}
\email{guoda@corp.netease.com}

\author{Shijia Wang}
% \authornote{Corresponding author.}
\affiliation{%
  \institution{NetEase Cloud Music}
  \city{Hangzhou}
  \country{China}}
\email{wangshijia1@corp.netease.com}

\author{Qiang Xiao}
\authornote{Corresponding author.}
\affiliation{%
  \institution{NetEase Cloud Music}
  \city{Hangzhou}
  \country{China}}
\email{hzxiaoqiang@corp.netease.com}

\author{Yintao Ren}
\affiliation{%
  \institution{NetEase Cloud Music}
  \city{Hangzhou}
  \country{China}}
\email{renyintao@corp.netease.com}

\author{Weisheng Li}
\affiliation{%
  \institution{NetEase Cloud Music}
  \city{Hangzhou}
  \country{China}}
\email{liweisheng01@corp.netease.com}

\author{Songpei Xu}
\affiliation{%
  \institution{NetEase Cloud Music}
  \city{Hangzhou}
  \country{China}}
\email{xusongpei@corp.netease.com}

\author{Ming Yue}
\affiliation{%
  \institution{NetEase Cloud Music}
  \city{Hangzhou}
  \country{China}}
\email{yueming03@corp.netease.com}

\author{Bin Huang}
\affiliation{%
  \institution{NetEase Cloud Music}
  \city{Hangzhou}
  \country{China}}
\email{huangbin02@corp.netease.com}

\author{Guanlin Wu}
\affiliation{%
  \institution{NetEase Cloud Music}
  \city{Hangzhou}
  \country{China}}
\email{wuguanlin03@corp.netease.com}

\author{Chuanjiang Luo}
\affiliation{%
  \institution{NetEase Cloud Music}
  \city{Hangzhou}
  \country{China}}
\email{luochuanjiang03@corp.netease.com}

%%
%% By default, the full list of authors will be used in the page
%% headers. Often, this list is too long, and will overlap
%% other information printed in the page headers. This command allows
%% the author to define a more concise list
%% of authors' names for this purpose.
\renewcommand{\shortauthors}{Guo et al.}

%%
%% The abstract is a short summary of the work to be presented in the
%% article.
\begin{abstract}
Generative retrieval has emerged as a promising paradigm in recommender systems, offering superior sequence modeling capabilities over traditional dual-tower architectures.
However, in large-scale industrial scenarios, such models often suffer from \textit{inherent myopia}: due to single-step inference and strict latency constraints, they tend to collapse diverse user intents into locally optimal predictions, failing to capture long-horizon and multi-item consumption patterns. Moreover, real-world retrieval systems must follow explicit retrieval instructions, such as category-level control and policy constraints. Incorporating such instruction-following behavior into generative retrieval remains challenging, as existing conditioning or post-hoc filtering approaches often compromise relevance or efficiency.
In this work, we present \text{\CLIMBER}, a unified generative retrieval framework to address both limitations.
First, we introduce Time-Aware Multi-Item Prediction (TAMIP), a novel training paradigm designed to mitigate inherent myopia in generative retrieval. By distilling long-horizon, multi-item foresight into model parameters through time-aware masking, TAMIP alleviates locally optimal predictions while preserving efficient single-step inference.
Second, to support flexible instruction-following retrieval, we propose Condition-Guided Sparse Attention (CGSA), which incorporates business constraints directly into the generative process via sparse attention, without introducing additional inference steps.
Extensive offline experiments and online A/B testing at NetEase Cloud Music, one of the largest music streaming platforms, demonstrate that Climber-Pilot consistently outperforms state-of-the-art baselines, achieving a 4.24\% lift of the core business metric.
\end{abstract}

%%
%% The code below is generated by the tool at http://dl.acm.org/ccs.cfm.
%% Please copy and paste the code instead of the example below.
%%


%%
%% Keywords. The author(s) should pick words that accurately describe
%% the work being presented. Separate the keywords with commas.
%% A "teaser" image appears between the author and affiliation
%% information and the body of the document, and typically spans the
%% page.

%%
%% This command processes the author and affiliation and title
%% information and builds the first part of the formatted document.
\maketitle

\section{Introduction}

Large-scale recommender systems rely on retrieval models to efficiently narrow down a massive item corpus into a small set of candidates for downstream ranking. In industrial settings, retrieval is required to operate under strict latency constraints while serving billions of requests per day, making both efficiency and robustness first-order concerns. Traditional retrieval systems are predominantly built upon dual-tower architectures~\cite{covington2016deep,shen2014learning,yi2019sampling}, which independently encode users and items and perform candidate selection via vector similarity search. While computationally efficient, such architectures fundamentally limit the expressive capacity of retrieval models, as rich and heterogeneous user interaction histories must be compressed into a single static representation. Although recent industrial efforts have explored multimodal feature fusion~\cite{yang2024cascading}, multi-task learning~\cite{wang2024ppen}, and cross-domain interest transfer~\cite{pan2024usit} to enrich representations within this paradigm, the underlying expressivity bottleneck remains.

Generative retrieval~\cite{rajput2023recommender,hou2025generating,ju2025generative} has recently emerged as a promising alternative paradigm, redefining candidate generation as a conditional sequence modeling problem. By autoregressively modeling user interaction sequences, generative retrievers are able to capture fine-grained temporal dependencies and complex behavioral patterns that are difficult to express in embedding-based retrieval frameworks. Moreover, generative approaches enable end-to-end optimization without relying on approximate nearest-neighbor search during inference, offering a unified modeling framework for retrieval.

Despite their advantages, generative retrieval models face significant challenges when deployed in large-scale industrial systems. One fundamental limitation is what we refer to as \textit{inherent myopia}~\cite{feng2024long,liu2025mtp,huang2025listwise}. In practice, latency constraints typically restrict generative retrievers to single-step inference at serving time. Consequently, most existing approaches are trained with next-item prediction objectives, encouraging the model to optimize for immediate relevance rather than long-horizon user intent~\cite{ie2019slateq}. This training-inference mismatch often leads to locally optimal predictions, where diverse future interests are collapsed into a narrow set of highly probable items, limiting both exploration and coverage.

An equally critical challenge arises from the operational requirements of industrial retrieval systems. Beyond relevance, retrieval models are often required to follow explicit retrieval instructions during candidate generation~\cite{yan2025lum,agarwal2025pinrec}, such as category-level constraints, policy rules, or business-driven control signals. In traditional retrieval pipelines, these requirements are typically addressed through heuristic filtering or rule-based post-processing. However, such approaches are poorly suited to generative retrieval models. Naively conditioning generation on control signals often provides insufficient precision, while post-hoc filtering can degrade relevance or introduce additional inference overhead, undermining the efficiency advantages of generative retrieval.

Addressing these challenges requires rethinking where and how complexity is handled in generative retrieval systems. Rather than introducing multi-step inference or relying on pipeline-level heuristics, we argue that the key lies in shifting complexity from inference to training, and from post-hoc constraint handling to model-internal inductive biases. Following this principle, we propose \text{\CLIMBER}, a unified generative retrieval framework designed for large-scale industrial deployment. \text{\CLIMBER} enables generative retrievers to capture long-horizon, multi-item user intent~\cite{donkers2017sequential} and to follow explicit retrieval instructions, while preserving the efficiency of single-step inference.

\text{\CLIMBER} achieves this through two complementary design choices. First, it distills long-horizon, multi-item foresight into model parameters during training by explicitly modeling batch-based exposure and delayed consumption patterns. This training-time distillation mitigates the myopic behavior induced by single-step objectives, allowing the model to internalize future-oriented signals without increasing inference complexity. Second, \text{\CLIMBER} embeds retrieval instructions directly into the generative process through attention-level control, enabling constraints to influence candidate generation in a fine-grained and efficient manner, without resorting to post-hoc filtering.

We evaluate \text{\CLIMBER} in a large-scale industrial recommendation system through extensive offline experiments and online A/B testing. The results demonstrate that \text{\CLIMBER} consistently outperforms state-of-the-art generative retrieval baselines in both retrieval quality and controllability, delivering consistent improvements in core business metrics under strict latency constraints. Our contributions can be summarized as follows:
\begin{itemize}[leftmargin=1.8em, itemsep=0.2em, topsep=2pt]
\item We identify inherent myopia as a fundamental limitation of generative retrieval under single-step inference, and reveal that consumption lag in batch exposure settings further exacerbates this issue in industrial systems.
\item We propose a training-time distillation paradigm that enables generative retrievers to internalize long-horizon, multi-item user intent without increasing inference complexity.
\item We introduce an attention-level conditioning framework that supports explicit instruction-following during generation, while allowing delayed consumption signals to be incorporated into the retrieval process.
\item We validate the proposed framework through extensive offline experiments and large-scale online deployment in a real-world production system.
\end{itemize}

\section{Related Work}

Sequential recommendation models, exemplified by SASRec~\cite{kang2018sasrec} and BERT4Rec~\cite{sun2019bert4rec}, have established the standard for capturing long-range dependencies in user behavior via self-attention mechanisms. Consequently, the paradigm has shifted toward generative recommendation frameworks such as P5~\cite{geng2022P5} and VQ-Rec~\cite{hou2023VQRec}, which reformulate retrieval as a sequence-to-sequence generation task. Industrial adaptations like OneRec~\cite{deng2025onerec} and HSTU~\cite{zhai2024hstu} have further demonstrated the scalability of this approach in high-throughput environments. Beyond architectural advances, recent works have also refined semantic identifier design through progressive residual quantization~\cite{wang2025psrq,xiao2026beyond} and hierarchical structure-aware multimodal modeling~\cite{pan2026hisam}, providing complementary signals for generative retrieval.

Despite these advancements, generative retrieval models predominantly rely on the Next Item Prediction (NIP) objective. Driven by the greedy nature of NIP optimization, model predictions tend to collapse toward high-frequency "next" items, which leads to issues with diversity and long-term engagement. Although traditional multi-interest models like MIND~\cite{li2019mind} and ComiRec~\cite{cen2020comirec} address diversity through static capsule routing, they lack the capacity to model dynamic interest trajectories within a generative framework.

LUM~\cite{yan2025lum} and PinRec~\cite{agarwal2025pinrec} used the in-context learning capabilities of generative architectures to explore conditioned retrieval. However, their approaches are primarily limited to adjusting retrieval results based on coarse-grained features, such as action types. Crucially, they fail to incorporate the diverse and complex business requirements—such as specific genres, languages, or item freshness—directly into the retrieval model. This limitation indicates a lack of genuine Instruction Following capability, significantly restricting their applicability in sophisticated industrial scenarios where flexible, policy-aware recommendation is imperative.

\section{Method}

\begin{figure*}[h]
  \centering
  \includegraphics[width=0.95\textwidth]{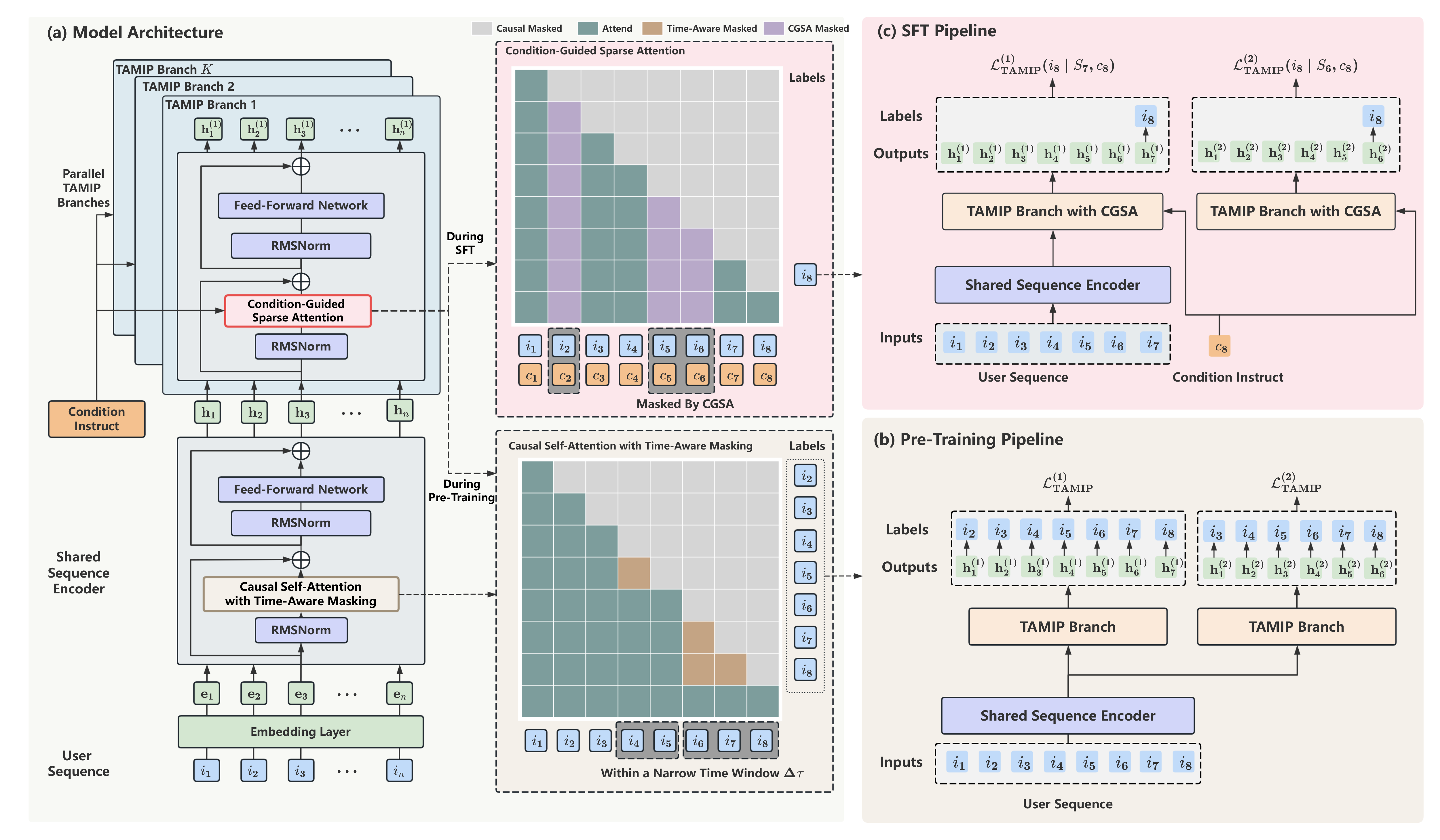}
  \caption{Overview of the \text{\CLIMBER} framework. (a) Model Architecture. This part details the Time-Aware mask employed in the TAMIP module and illustrates the working mechanism of CGSA. During the SFT stage, the TAMIP branch adopts CGSA to enable instruction-following capability. (b) Pre-Training Pipeline. (c) SFT Pipeline.}
  \Description{Overall Framework of the Climber-Pilot model, including Time-Aware Multi-Item Prediction architecture and Condition-Guided Sparse Attention mechanism.}
  \label{fig:frame}
\end{figure*}

\subsection{Framework Overview}
In this section, we propose a novel sequential recommendation framework, \textbf{\CLIMBER}, which follows a two-stage pre-training and fine-tuning paradigm. During the pre-training stage, the model learns to capture future interest representations of users via a next-item prediction objective. To alleviate the myopia issue inherent in generative architectures, we propose the Time-Aware Multi-Item Prediction method. Subsequently, in the fine-tuning stage, the model is optimized using a conditioned next-item prediction approach. By introducing conditional signals (e.g., music genres), we guide the model to learn interest representations within fine-grained subcategories, thereby enabling targeted recall. The overall training pipeline is illustrated in Figure~\ref{fig:frame}. 

\subsection{General Retrieval Capability Pre-Training}
\label{sec:pre-training}
\subsubsection{Sequence Encoding}
Let $S_n = \{i_1, i_2, \dots, i_n\}$ denote the historical interaction sequence of user $u$, where $i_k \in \mathcal{I}$ represents the $k$-th item interacted with by the user. To derive a comprehensive representation for each item $i_k$, we concatenate a learnable ID embedding with its corresponding category feature embeddings. We utilize a hash-based embedding table for ID storage. Consequently, the raw interaction sequence $S_n$ is transformed into a dense embedding sequence: $\mathbf{E}_n = \{\mathbf{e}_1, \mathbf{e}_2, \dots, \mathbf{e}_n\} \in \mathbb{R}^{n \times d}$,
where $\mathbf{e}_k$ denotes the fused embedding vector of item $i_k$ and $d$ is the embedding dimension.

% \subsubsection{Transformer Encoder}
% To capture complex autoregressive dependencies within the user's behavior sequence, we employ a stack of $L$ causal Transformer blocks as the backbone encoder. 

The embedding sequence $\mathbf{E}_n$ is fed into the encoder. Following \cite{xu2025climber}, we employ a multi-layer Transformer architecture incorporating adjusted relative attention bias and a causal mask. This design prevents information leakage from future positions and yields a sequence of contextualized hidden states $\mathbf{H}_n = \{\mathbf{h}_1, \dots, \mathbf{h}_n\} \in \mathbb{R}^{n \times d}$.

\subsubsection{Time-Aware Multi-Item Prediction}
\label{sec:time_aware_mip}

The standard next item prediction paradigm suffers from an Inherent Myopia problem~\cite{liu2025mtp}. 
In the domain of NLP, a prevalent strategy to mitigate such myopia is to extend the prediction horizon via Multi-Token Prediction (MTP)~\cite{gloeckle2024better} objectives.
However, directly applying multi-item objectives in industrial settings is challenging due to a misalignment between batch serving and sequential logging. As noted by~\cite{liang2025tbgrecall}, items within the same retrieval request have no intra-request causal dependencies—causality exists only across requests. Yet interaction logs mechanically serialize these co-exposed items into a dense sequence. 
Formally, for a single request $r$, $m$ items are exposed to the user simultaneously, yet interaction logs record them as a serialized sequence $\{i_1, i_2, \dots, i_m\}$, imposing an artificial ordering that does not reflect true user intent progression.
We term this phenomenon \textbf{\textit{Consumption Lag}}: models trained on such logs learn spurious sequential patterns from what is fundamentally a parallel process.
Consequently, the model overfits to artificial within-batch patterns. These observations indicate that standard sequence modeling is insufficient for batch-based retrieval.

To fundamentally alleviate this limitation, we introduce Time-Aware Multi-Item Prediction. Structurally, TAMIP is composed of two integral parts: a multi-branch prediction backbone designed for expanded foresight, and a pivotal Time-Aware Masking mechanism that serves as the core improvement to filter spurious correlations.
Formally, to extend the model's foresight, we construct $K$ parallel projection branches on top of the shared user representation $\mathbf{h}_n$. Each branch is implemented as a single Transformer layer with independent (non-shared) parameters, transforming the shared input $\mathbf{h}_n$ into a step-specific representation $\mathbf{h}^{(k)}_n$, responsible for predicting the item $i_{n+k}$. In this setting, the case of $K=1$ aligns with the standard NIP objective.

However, the naive multi-step prediction is vulnerable to the aforementioned consumption lag. To eliminate this contamination and force the model to learn genuine long-term dependencies, we introduce the Time-Aware Masking mechanism (illustrated in Figure~\ref{fig:frame}(a)). 

To align with the batch serving mechanism used in online production environments, our model predicts the target sequence $\{i_{n+1}, \dots, i_{n+K}\}$ in a single pass. Consequently, the temporal visibility for all prediction heads is anchored to the timestamp of the first target item, $\tau_{n+1}$. Therefore, we impose a shared temporal constraint on both the backbone encoder and all TAMIP prediction branches. We define a safe margin $\Delta \tau$, empirically set to the average of logged serving time intervals, amounting to 15 minutes in our production system. Specifically, we explicitly mask any historical interaction $i_j$ whose timestamp $\tau_j$ falls within the interval $[\tau_{n+1} - \Delta \tau, \tau_{n+1}]$. Formally, given a user sequence $S$ with timestamps $\tau$, the time-aware attention mask $M_{\text{temp}}\in \mathbb{R}^{n\times n}$ is defined as:

\begin{equation}
M_{\text{temp}}(i,j) = 
\begin{cases} 
-\infty & \text{if } \tau_j \in [\tau_{n+1} - \Delta \tau, \tau_{n+1}] \\
0 & \text{otherwise.}
\end{cases}
\end{equation}

This mask is integrated into the self-attention mechanism alongside the standard causal mask $M_{\text{causal}}$, which forbids positions from attending to future tokens (i.e., $M_{\text{causal}}(i, j) = -\infty$ if $j > i$ and $0$ otherwise). Let $Q, K, V \in \mathbb{R}^{n \times d_k}$ denote the query, key, and value matrices linearly projected from the hidden states, where $d_k$ is the projection dimension. The final attention computation is formulated as:
\begin{equation}
\text{Attention}(Q, K, V) = \text{softmax}\left(\frac{QK^\top}{\sqrt{d_k}} + M_{\text{causal}} + M_{\text{temp}}\right)V.
\end{equation}

Note that employing $\tau_{n+1}$ in $M_{\text{temp}}$ does not introduce data leakage: this timestamp is solely used to mask out recent interactions, thereby simulating the information horizon available at inference time. No label-specific features beyond the temporal cutoff are exposed to the model. In addition, the time-aware mask is only used during training. At serving time, recent interactions are not yet logged due to consumption lag, so the same information gap arises naturally and no timestamp is needed.

By strictly prohibiting information flow from temporally proximal tokens, TAMIP prevents the model from relying on short-term behavioral bursts or batch-induced artifacts. Instead, it compels the model to infer future intents based on historical context, thereby effectively alleviating inherent myopia and enhancing the model's ability to follow long-term user interest trajectories.

\subsubsection{TAMIP Loss Formulation}
For each TAMIP branch $k \in \{1, \dots, K\}$, we predict the item $i_{n+k}$ conditioned on the augmented representation $\mathbf{h}^{(k)}_n \in \mathbb{R}^d$. 
Since computing the full softmax over a massive item space is computationally intractable,  we adopt a sampled softmax strategy~\cite{pancha2022pinnerformer} shared across all prediction heads. The loss for the $k$-th prediction head is defined as:
\begin{equation}
\mathcal{L}_{\text{TAMIP}}^{(k)}(i_{n+k} \mid S_n) = - \log \frac{\exp(\phi(\mathbf{h}^{(k)}_n, \mathbf{e}_{n+k}))}{\exp(\phi(\mathbf{h}^{(k)}_n, \mathbf{e}_{n+k})) + \sum_{j \in \mathcal{N}} \exp(\phi(\mathbf{h}^{(k)}_n, \mathbf{e}_j))},
\end{equation}
where $S_n = \{i_1, \dots, i_{n}\}$ represents the user interaction history, $\phi(\mathbf{a}, \mathbf{b}) = \mathbf{a}^\top \mathbf{b}$ denotes the inner product scoring function, $\mathbf{e}_{n+k} \in \mathbb{R}^d$ is the embedding of the target item at step $n+k$, and $\mathcal{N}$ represents the shared in-batch negative set.

The overall pre-training objective $\mathcal{L}_{\text{PT}}$ aggregates contributions from all $K$ TAMIP branches:
\begin{equation}
\mathcal{L}_{\text{PT}} = \sum_{k=1}^{K}  \mathcal{L}_{\text{TAMIP}}^{(k)}.
\end{equation}
By optimizing this objective on massive user behavior corpora, we conduct a pre-training stage that endows the model with general retrieval capabilities. Crucially, the proposed TAMIP strategy effectively alleviates the inherent myopia of generative architectures by enforcing multi-step foresight.

\subsection{Instruction Following via SFT}
\label{sec:sft}

While TAMIP-based pre-training endows the model with the capability to anticipate future user interests, the lack of explicit instruction-following capabilities hinders the \text{\CLIMBER} model from addressing the diverse and dynamic requirements of industrial systems. Drawing upon the SFT paradigm from LLMs~\cite{gunel2020supervised}, which has been further extended to personalized recommendation tasks~\cite{pan2026l2rec,wang2025lemon}, we conceptualize specific industrial retrieval conditions—such as genres, languages, new releases, or classics—as the equivalent of instructions. Consequently, we reframe these conditional retrieval requirements as distinct instruction-following tasks. To achieve this, we construct condition-specific SFT datasets and propose a novel Condition-Guided Sparse Attention mechanism. This process effectively transforms the model from an unconditional generator into a controllable and instruction-aware retriever.

By conditioning the model on explicit prompts, we enable dynamic, task-specific candidate generation. This design offers a significant efficiency advantage: a single unified model can now replace multiple specialized retrieval pipelines simply by varying the prompt. Consequently, this consolidation drastically reduces deployment complexity and maintenance costs.

\subsubsection{Construction of Condition-Specific Retrieval Datasets}
\label{subsubsec:sft-datasets}
We construct the SFT dataset by collecting high-quality online user interaction logs that align with specific retrieval conditions. These conditions encompass diverse scenarios, such as retrieving items from the "New Songs" candidates or targeting items with specific attributes (e.g., the Rock tag). Formally, we denote the set of these retrieval conditions as $\mathcal{C}$. For each target item in the SFT dataset, its corresponding condition is denoted as $c \in \mathcal{C}$.

First, we formulate the task as an instruction-following problem. In this setup, the retrieval condition $c$ serves as an explicit instruction, guiding the model to predict the target item $i_{n+1}$ based on the user sequence $\{i_1, \dots, i_n\}$. Note that the task remains non-trivial due to the large cardinality of the item pool associated with each condition, effectively mitigating data leakage risks.

Second, we repurpose TAMIP's multi-branch structure in SFT. In the pre-training phase, TAMIP uses a single context to predict a future trajectory ($i_{n+1}, \dots, i_{n+K}$). In the SFT phase, however, we shift this paradigm: we utilize multiple truncated context windows—specifically $\{i_1, \dots, i_{n-K+1}\}$ through $\{i_1, \dots, i_n\}$—as inputs for separate branches. Each branch, conditioned on its specific context length and $c_{n+1}$ (the retrieval condition associated with $i_{n+1}$), is trained to predict the identical target item $i_{n+1}$. As illustrated in Figure~\ref{fig:frame}(c), this design enables the reuse of a single instruction condition $c_{n+1}$ across all $K$ TAMIP branches, significantly improving computational efficiency during training. Beyond efficiency, exposing the model to truncated contexts of different lengths forces it to follow the same instruction under different amounts of history. This improves robustness, since the history length seen at serving time also varies. Meanwhile, since each branch processes contexts of varying lengths, the model retains its capacity to capture multi-horizon dependencies—preserving the myopia-alleviating benefits established during pre-training.

\subsubsection{Condition-Guided Sparse Attention Mechanism}

User behaviors are inherently multimodal and noisy. Standard Multi-Head Attention (MHA) aggregates signals globally. This often accumulates noise from irrelevant categories, diluting the specific intent required for retrieval. To address this, we propose the Condition-Guided Sparse Attention (CGSA) mechanism. Unlike standard MHA, CGSA utilizes the instruction $c$ to actively denoise the context. It steers the model to focus exclusively on historical behaviors semantically aligned with $c$.

To enforce this compliance, we construct a sparse mask $M_{\text{sparse}} \in \mathbb{R}^{n\times n}$ that restricts attention connections to relevant items. Formally, for a sequence of length $n$, the attention mask at position $(i, j)$ is defined as:

\begin{equation}
M_{\text{sparse}}(i,j) = 
\begin{cases} 
0, & \text{if } \mathcal{C}(\text{item}_j) = c \\
% \quad \text{(Instruction Alignment)} \\
-\infty, & \text{otherwise,}
\end{cases}
\end{equation}

where $\mathcal{C}(\cdot)$ maps an item to its category space, and $c$ represents the target instruction condition.

The CGSA mechanism explicitly filters the long-term user history by zeroing out attention weights on items that do not match the target instruction, thereby denoising the behavioral sequence and focusing the model on condition-relevant interactions. 

We apply CGSA exclusively to the final TAMIP branches while retaining the same attention mask configuration as in the pre-training stage for the first $L$ Transformer layers. This choice is primarily motivated by computational efficiency: the shared $L$-layer backbone can be computed once and reused across multiple condition-specific branches, enabling efficient batched multi-condition inference. Another benefit of reusing the pre-training attention mask in the first $L$ layers is that it preserves the rich semantic representations learned during pre-training.

\subsubsection{SFT Optimization}

Initializing with pre-trained weights allows the model to retain general retrieval capabilities, and the objective evolves from unconditional prediction to instruction-conditioned retrieval.

In this stage, the proposed model is optimized to retrieve the ground-truth item $i_{n+1}$ constrained by condition $c_{n+1}$. Crucially, the CGSA mechanism is activated during this phase, filtering out noise and steering the TAMIP branches to focus on instruction-relevant items of users' interaction sequence.

To mitigate catastrophic forgetting of the general preferences learned during pre-training, a reduced learning rate was adopted. The SFT optimization specifically targets the multi-branch prediction capability. Specifically, for branch $k=1$, the model uses the full history $S_n$; for branch $k=K$, it uses the shortest history $S_{n-K+1}$. Formally, the loss function is defined as:
\begin{equation}
    \mathcal{L}_{\text{SFT}} =  \sum_{k=1}^{K} \mathcal{L}_{\text{TAMIP}}^{(k)}(i_{n+1} \mid S_{n-k+1}, c_{n+1}),
\end{equation}

where $S_{n-k+1}$ represents the truncated history window for the $k$-th TAMIP branch,  $c_{n+1}$ denotes the target instruction and $K$ is the number of TAMIP branches. This objective enforces the model to align its predictions with the explicit condition $c_{n+1}$.

\subsection{Deployment Detail}
\label{sec:inference}

\begin{figure}[h]
  \centering
  \includegraphics[width=0.95\linewidth]{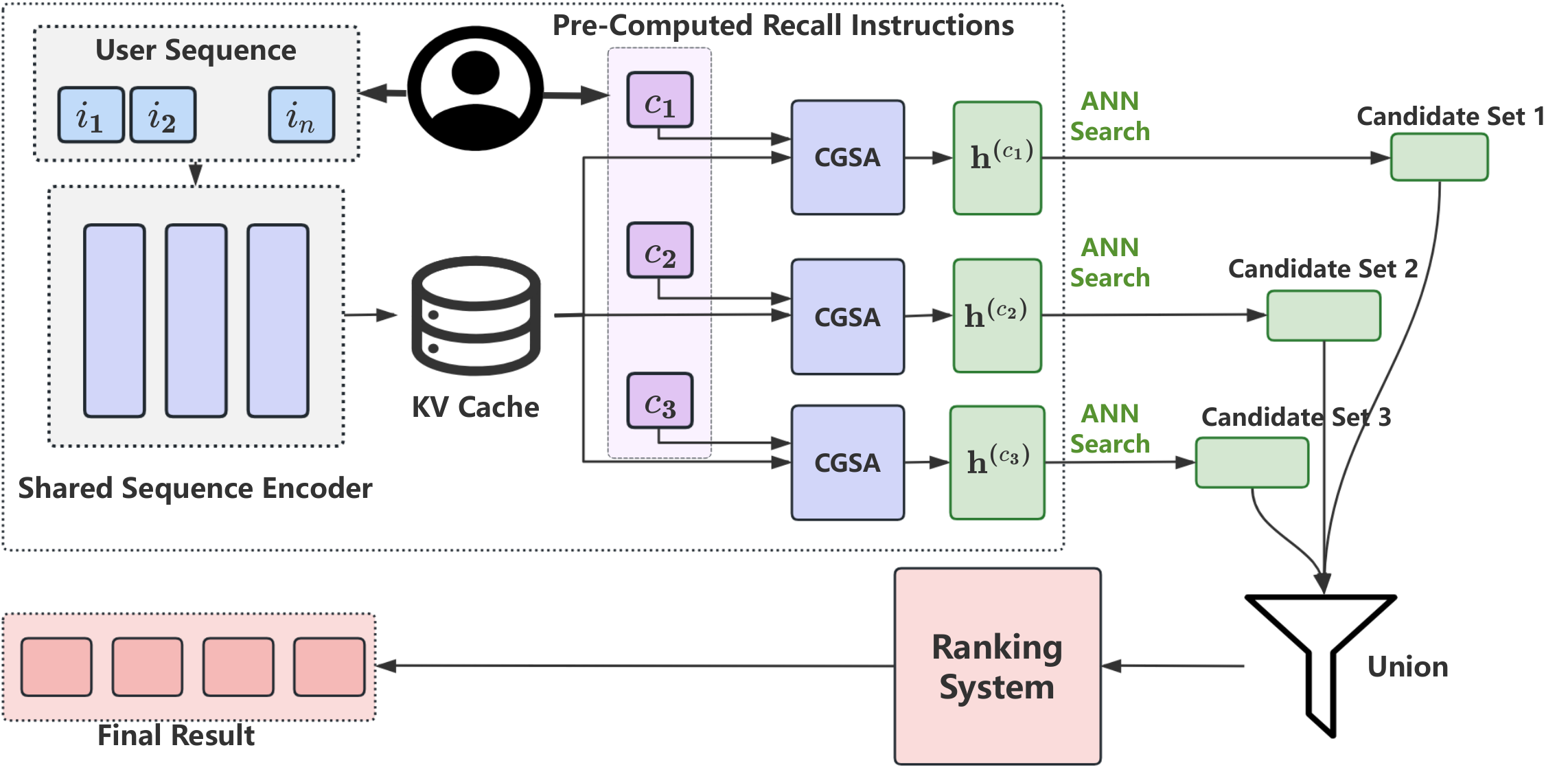}
  \caption{Illustration of the \text{\CLIMBER} inference process. 
  % The model uses multiple instructions to retrieve diverse candidates in a single request, achieving the capability of a multi-channel retrieval system within a single unified model.}
  }
  \Description{Inference pipeline of Climber-Pilot showing batched multi-condition retrieval with pre-computed recall instructions.}
  \label{fig:infer}
\end{figure}

The deployment details of \text{\CLIMBER} are illustrated in Figure~\ref{fig:infer}. We train our model on a cluster of 8$\times$A100 GPUs using a behavioral corpus of 45B interactions, and deploy the inference service via the TensorRT~\cite{davoodi2019tensorrt} framework. When a user initiates a request, the system synchronously fetches the behavior sequence and retrieves the cached recall instructions $\{c_1, \dots, c_P\}$, where $P$ denotes the number of pre-computed instructions.

\subsubsection{Pre-Computed Recall Instructions}
\label{subsubsec:pre-computed}
To enable efficient instruction-following inference, we pre-compute the top-$P$ interest categories for each user based on their recent interaction history. Specifically, we aggregate the user's behavioral signals over a sliding window and identify the most frequently engaged categories (e.g., Hip-Hop, New Releases, English songs). These category labels are then converted into recall instructions $\{c_1, \dots, c_P\}$ and cached for real-time retrieval. This design decouples the instruction generation from the online serving path, ensuring that the model can dynamically adapt to user preferences without incurring additional latency during inference.

\subsubsection{Batched Multi-Condition Inference}
Inspired by Climber~\cite{xu2025climber}, we propose Batched Multi-Condition Inference to achieve high throughput under strict latency constraints. Since the CGSA mechanism affects only the final branches, we can bundle all $P$ conditions into a single inference pass. 

The computation proceeds in two steps. First, the initial $L$ shared Transformer blocks process the user sequence only once, and the resulting intermediate representations are cached. Subsequently, these cached representations are reused across the final layer for all $P$ condition branches. This design effectively avoids redundant computation for the heavy backbone network.

\section{Experiments}
% 需要按 RQ 展开吗
\subsection{Experimental Setup}

\subsubsection{Datasets}

\begin{table}[h]
\centering
\caption{Statistics of the processed datasets. ``Avg. $n$'' denotes the average sequence length per user.}
\label{tab:dataset_statistics}
\begin{tabular}{lccc}
\toprule
Dataset & \#Users & \#Items & Avg. $n$ \\
\midrule
Amazon-Sports & 18,357 & 35,598 & 8.32 \\
Amazon-Beauty & 22,363 & 12,101 & 8.87 \\
Amazon-Toys   & 19,412 & 11,924 & 8.63 \\
Industrial    & $>$40M & $>$6M  & $>$100 \\
\bottomrule
\end{tabular}
\end{table}

We conduct experiments on two large-scale datasets: (1) Amazon Benchmarks, including Amazon-Sports, Amazon-Beauty, and Amazon-Toys. (2) Industrial, an industrial-scale dataset derived from real-world user interaction logs of the NetEase Cloud Music platform. Table~\ref{tab:dataset_statistics} summarizes the main statistics for both datasets.

To evaluate the instruction-following capability of \text{\CLIMBER}, we further curate a dedicated SFT dataset from online logs that does not overlap with the pre-training corpus. In all offline experiments regarding controllable generation, we adopt music genre and language as representative conditional signals.

\subsubsection{Implementation Details and Metrics}

\paragraph{Model Architecture.}
The proposed model is instantiated using a 3-layer Transformer~\cite{vaswani2017attention} backbone. We set the embedding dimension to $d=128$ for the industrial dataset and $d=32$ for the Amazon datasets. Since additional TAMIP branches introduce extra computational overhead during both training and inference, we set $K=2$ to balance efficiency and the core objective of validating TAMIP's effectiveness in alleviating myopia.

\paragraph{Training Configuration.}
All models use the Adam optimizer~\cite{kingma2014adam} with a learning rate of $5 \times 10^{-4}$ and a weight decay of $1 \times 10^{-6}$. The batch size is fixed to 256. For the SFT stage, we adopt a reduced learning rate of $1 \times 10^{-5}$ to mitigate catastrophic forgetting of pre-trained knowledge. The time-aware masking margin $\Delta \tau$ is set to 15 minutes for the industrial dataset, corresponding to the average logged request interval.

\paragraph{Evaluation Metrics.}
To assess the retrieval quality during the candidate generation phase, we adopt Hit Rate at $K$ (HR@$K$) as the primary metric, with $K \in \{10, 20, 50\}$. Specifically, we treat the last item of each sequence as the ground-truth label and utilize the preceding interaction history to generate the user interest representation. To simulate large-scale retrieval, we build an approximate nearest neighbor (ANN) index over the entire item corpus using FAISS~\cite{douze2025faiss} with the HNSW algorithm~\cite{malkov2018efficient}, and retrieve the top-$K$ candidates. Formally, HR@$K$ is defined as:
\begin{equation}
    \text{HR@}K = \frac{1}{|\mathcal{D}|}\sum_{i=1}^{|\mathcal{D}|} \mathbb{I}[\text{rank}(y_i) \leq K],
\end{equation}
where $\mathcal{D}$ denotes the test set, $y_i$ represents the ground-truth item for the $i$-th instance, and $\mathbb{I}[\cdot]$ is the indicator function.

\subsection{Pre-training Evaluation: General Retrieval via TAMIP}
% Please add the following required packages to your document preamble:
% \usepackage{booktabs}
% \usepackage{multirow}
% Please add the following required packages to your document preamble:
% \usepackage{booktabs}
% \usepackage{multirow}
\begin{table*}[t]
\caption{Pre-training evaluation: overall performance comparison on multiple datasets. Best results are in bold, second-best are underlined.}
\label{tab:overall_performance}
\begin{tabular}{@{}c|ccc|ccc|ccc|ccc@{}}
\toprule
\multirow{2}{*}{Model} &
  \multicolumn{3}{c|}{Industrial} &
  \multicolumn{3}{c|}{Amazon-Sports} &
  \multicolumn{3}{c|}{Amazon-Beauty} &
  \multicolumn{3}{c}{Amazon-Toys} \\ \cmidrule(l){2-13} 
       & HR@50  & HR@20  & HR@10  & HR@50  & HR@20  & HR@10  & HR@50  & HR@20  & HR@10  & HR@50  & HR@20  & HR@10  \\ \midrule
SASRec & 7.62\% & 4.41\% & 2.77\% & 4.67\% & 2.98\% & 1.98\% & 9.35\% & 6.21\% & 4.44\% & 8.42\% & 5.79\% & 4.51\% \\
TIGER  & 8.34\% & 5.15\% & \underline{3.82\%} & \underline{7.14\%} & \underline{4.40\%} & \underline{3.05\%} & 11.43\% & 8.39\% & 6.10\% & \underline{10.33\%} & \underline{7.11\%} & \underline{4.97\%} \\
HSTU   & \underline{9.50\%} & \underline{5.31\%} & 3.24\% & 6.49\% & 4.31\% & 2.83\% & 7.43\% & 4.96\% & 3.46\% & 7.83\% & 5.10\% & 3.63\% \\
PinRec & 8.78\% & 5.00\% & 3.07\% & 6.63\% & 4.30\% & 2.98\% & \underline{12.08\%} & \underline{8.53\%} & \underline{6.32\%} & 8.52\% & 5.15\% & 3.66\% \\
\textbf{\text{\CLIMBER}} &
  \textbf{10.95\%} &
  \textbf{6.40\%} &
  \textbf{4.02\%} &
  \textbf{7.20\%} &
  \textbf{4.62\%} &
  \textbf{3.28\%} &
  \textbf{12.12\%} &
  \textbf{8.95\%} &
  \textbf{6.74\%} &
  \textbf{11.20\%} &
  \textbf{8.01\%} &
  \textbf{6.13\%} \\ \bottomrule
\end{tabular}
\end{table*}

A comprehensive comparison was conducted between \text{\CLIMBER} and several SOTA models, including both sequential recommendation and generative retrieval baselines, on an industrial dataset and three public Amazon datasets. To ensure a fair comparison with standard baselines, we evaluated the pre-trained \text{\CLIMBER} utilizing solely its primary TAMIP branch (targeting the immediate next item, i.e., $i_{n+1}$), excluding auxiliary long-term TAMIP branches used during training.

As presented in Table~\ref{tab:overall_performance}, \text{\CLIMBER} consistently demonstrates superior performance across all metrics. We attribute this significant improvement to the proposed TAMIP mechanism. While traditional SOTA models rely on the standard NIP paradigm, which often suffers from inherent myopia due to a narrow focus on immediate supervision, \text{\CLIMBER} effectively mitigates this issue by incorporating multi-interval supervision to learn robust user preferences.

\subsection{Effect of Time-Aware Masking}

% \begin{table*}
% \caption{Performance comparison}
% \label{tab:myopia_verification}
% \centering
% \begin{tabular}{clcccc}
% \toprule
% % 下面这一行末尾原本是单反斜杠 \，已改为双反斜杠 \\
% \textbf{Method} & \textbf{Step} & \textbf{HR@100} & \textbf{HR@50} & \textbf{HR@20} & \textbf{HR@10} \\
% \midrule
% % 建议将多行合并的第二个参数改为 *，表示自动宽度
% \multirow{10}{*}{\textbf{w/o MTP}}
% & 1 & 0.1117 & 0.0762 & 0.0441 & 0.0277 \\
% & 2 & 0.1006 & 0.0668 & 0.0367 & 0.0220 \\
% & 3 & 0.0959 & 0.0622 & 0.0331 & 0.0197 \\
% & 4 & 0.0931 & 0.0599 & 0.0317 & 0.0185 \\
% & 5 & 0.0905 & 0.0581 & 0.0304 & 0.0176 \\
% & 6 & 0.0880 & 0.0560 & 0.0291 & 0.0168 \\
% & 7 & 0.0875 & 0.0555 & 0.0286 & 0.0168 \\
% & 8 & 0.0859 & 0.0543 & 0.0279 & 0.0160 \\
% & 9 & 0.0846 & 0.0530 & 0.0271 & 0.0156 \\
% & 10 & 0.0850 & 0.0536 & 0.0272 & 0.0156 \\
% \midrule
% \multirow{10}{*}{\textbf{w/ Time-Aware MTP}}
% & 1 & 0.1562 & 0.1095 & 0.0640 & 0.0402 \\
% & 2 & 0.1500 & 0.1041 & 0.0605 & 0.0379 \\
% & 3 & 0.1486 & 0.1027 & 0.0593 & 0.0371 \\
% & 4 & 0.1475 & 0.1023 & 0.0589 & 0.0373 \\
% & 5 & 0.1476 & 0.1020 & 0.0588 & 0.0372 \\
% & 6 & 0.1481 & 0.1028 & 0.0596 & 0.0374 \\
% & 7 & 0.1607 & 0.1126 & 0.0660 & 0.0425 \\
% & 8 & 0.1585 & 0.1113 & 0.0653 & 0.0417 \\
% & 9 & 0.1495 & 0.1034 & 0.0603 & 0.0381 \\
% & 10 & 0.1488 & 0.1035 & 0.0603 & 0.0380 \\
% \bottomrule
% \end{tabular}
% \end{table*}

\subsubsection{Three-Way Comparison: NIP vs MIP vs TAMIP}

\begin{figure*}[h]
  \centering
  \includegraphics[width=0.9\textwidth]{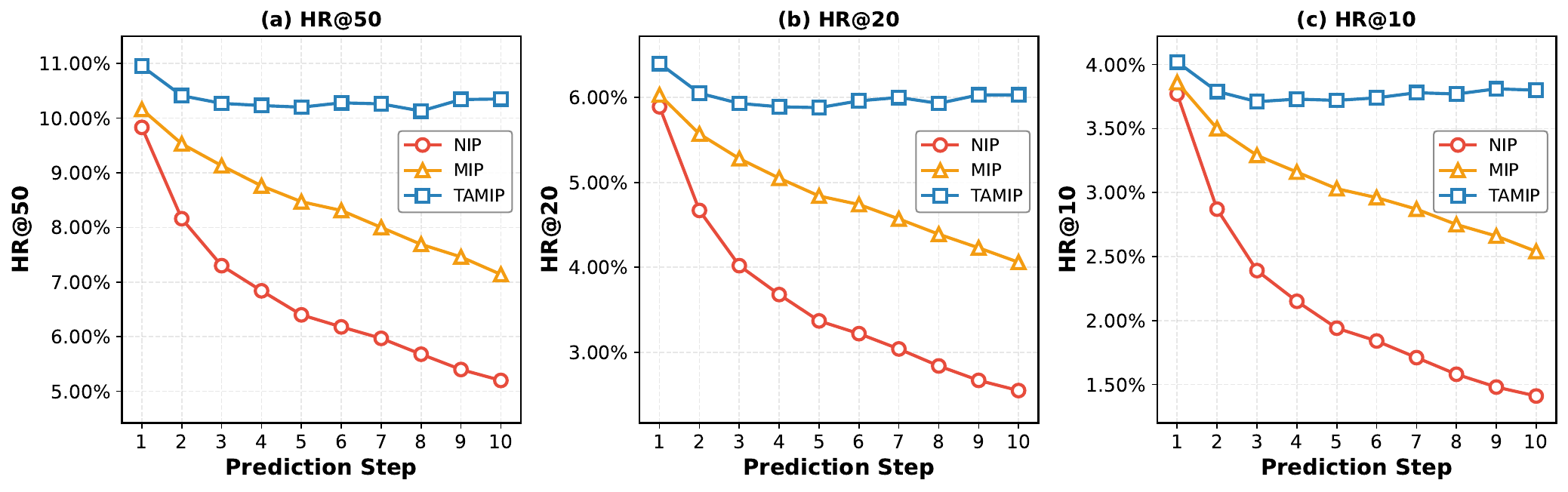}
  \caption{Effectiveness of TAMIP in alleviating inherent myopia. We compare three training paradigms on the industrial dataset: NIP (red), MIP (orange), and TAMIP (blue). The TAMIP curves demonstrate remarkable long-term stability.}
  \Description{Performance comparison showing myopia verification with and without TAMIP across multiple prediction horizons.}
  \label{fig:myopia}
\end{figure*}

We verify the effectiveness of TAMIP in alleviating inherent myopia by comparing it with NIP and MIP (Multi-Item Prediction without time-aware masking) on the \text{\CLIMBER} architecture. As shown in Figure~\ref{fig:myopia}, we use retrieval results at time step $t$ to predict user interests over the subsequent 10 steps ($t+1, \ldots, t+10$).

Two key findings emerge from the results. First, NIP exhibits a pronounced performance decline as the prediction horizon extends, confirming the presence of inherent myopia caused by its greedy, single-step optimization objective. MIP partially mitigates this issue with modest gains in initial-step HR@$K$ and a gentler downward trend, yet still suffers from noticeable long-term decay.

Second, TAMIP yields a dual advantage: it significantly elevates the overall performance baseline while maintaining remarkable stability. We attribute this improvement to the time-aware masking mechanism, which effectively filters out artificial within-batch patterns and forces the model to rely on genuine long-term behavioral signals.

\subsubsection{Sensitivity to $\Delta\tau$}

The above comparison shows the binary effect of applying time-aware masking. As noted in Sec.~\ref{sec:time_aware_mip}, the production setting uses $\Delta\tau = 15$ min based on the average logged serving-time interval. We now analyze the sensitivity of TAMIP to deviations from this value. Since $\Delta\tau = 0$ recovers the MIP variant, this study can also be viewed as a fine-grained ablation over the masking strength. We vary $\Delta\tau \in \{0, 7.5, 15, 30\}$ minutes on the industrial dataset. Results are shown in Table~\ref{tab:tau_sensitivity}.

\begin{table}[h]
\centering
\caption{Sensitivity of the time-aware masking window $\Delta\tau$ on the industrial dataset.}
\label{tab:tau_sensitivity}
\begin{tabular}{lccc}
\toprule
$\Delta\tau$ & HR@50 & HR@20 & HR@10 \\
\midrule
0 min   & 10.16\% & 6.03\% & 3.86\% \\
7.5 min & 10.11\% & 6.07\% & 3.91\% \\
15 min  & \textbf{10.95\%} & \textbf{6.40\%} & \textbf{4.02\%} \\
30 min  & 9.89\% & 5.94\% & 3.86\% \\
\bottomrule
\end{tabular}
\end{table}

The 15-minute setting is best across all metrics. From 0 to 7.5 min, all three metrics stay nearly flat. This indicates that a window narrower than the typical inter-request interval still leaves within-batch items visible to the model, so the spurious sequential patterns induced by consumption lag are not effectively suppressed. From 7.5 to 15 min, HR@50 rises from 10.11\% to 10.95\%, a relative gain of about 8\%, and HR@20 and HR@10 move in the same direction. The 15 min setting matches the system statistics already noted in Sec.~\ref{sec:time_aware_mip}: each request exposes 3--4 songs of about 4 minutes each, and 15 min is the average gap between batch consumption and the next request, which is the temporal scale at which consumption lag operates. From 15 to 30 min, performance drops back below the MIP variant ($\Delta\tau = 0$). A window much larger than this interval masks genuinely informative older interactions, which reduces the signal available for prediction and hurts model performance. The choice of $\Delta\tau$ is therefore tied to system-level statistics rather than free hyperparameter tuning.

\subsection{SFT Evaluation: Instruction Following via CGSA}

In this section, we evaluate the instruction-following retrieval capabilities of \text{\CLIMBER} on the industrial dataset. We configure the evaluation task as follows: given a user interaction sequence $S_n = \{i_1, \ldots, i_n\}$, we utilize the attribute (e.g., genre or language) of the ground-truth next item $i_{n+1}$ as the explicit instruction signal $c_{n+1}$. 

Our evaluation relies on two critical perspectives: (1) Recommendation Accuracy, measured by HR@$K$, and (2) Instruction Adherence, quantified by the Condition Compliance Rate.

Condition Compliance@$K$ (CC@$K$) formally quantifies the model's ability to translate explicit user intents or business constraints into the retrieved candidate set. Let $\mathcal{R}_K$ denote the set of top-$K$ items retrieved via the ANN procedure described above. We define CC@$K$ as the proportion of retrieved items that satisfy the specified condition $c$:
\begin{equation}
    \text{CC@}K = \frac{1}{K} \sum_{j \in \mathcal{R}_K} \mathbb{I}\left[ \mathcal{C}(j) = c \right],
\end{equation}
where $\mathcal{C}(j)$ represents the category associated with item $j$.

To validate the effectiveness of the proposed CGSA module, we compare our approach against three conditional generation paradigms: PinRec~\cite{agarwal2025pinrec}, LUM~\cite{yan2025lum}, two existing controllable retrieval models originally designed for coarse-grained action types, and AdaLN-Zero~\cite{peebles2023scalable}, a conditioning mechanism from Diffusion Transformers (DiT). For a fair comparison, all conditioning variants are fine-tuned from the same pre-trained checkpoint.

Table~\ref{tab:conditional_generation_comparison} summarizes the experimental results. We observe that the proposed CGSA achieves the highest performance across all HR@$K$ metrics, demonstrating superior recommendation accuracy. Regarding instruction adherence, although PinRec exhibits the highest CC, our proposed method maintains highly competitive performance.

While PinRec and LUM were designed for coarse-grained action-type conditioning, CGSA is tailored to the sequential modeling characteristics of recommendation. By filtering out irrelevant items from historical sequences, CGSA removes behavioral noise unrelated to the target condition. This yields two benefits: competitive CC values show that CGSA satisfies business constraints, while higher HR@$K$ indicates better alignment with user preferences. In practice, CGSA does not merely comply with instructions—it jointly optimizes for both business requirements and user satisfaction.

\begin{table*}[t]
\caption{SFT evaluation: performance comparison of conditional generation paradigms. The CGSA method demonstrates the highest HR@$K$ and robust performance in Condition Compliance.}
\label{tab:conditional_generation_comparison}
\centering
\begin{tabular}{lcccccc}
\toprule
\textbf{Model} & \textbf{HR@50} & \textbf{HR@20} & \textbf{HR@10} & \textbf{CC@50} & \textbf{CC@20} & \textbf{CC@10} \\
\midrule
PinRec & 13.82\% & 9.16\% & 6.29\% & \textbf{81.69\%} & \textbf{84.30\%} & \textbf{85.57\%} \\
LUM & 11.45\% & 7.51\% & 5.24\% & 68.97\% & 74.51\% & 77.84\% \\
AdaLN-Zero & \underline{15.84\%} & \underline{10.27\%} & \underline{6.88\%} & 75.49\% & \underline{80.41\%} & 80.96\% \\
\midrule
\textbf{CGSA} & \textbf{18.53\%} & \textbf{11.69\%} & \textbf{7.73\%} & \underline{78.58\%} & 80.07\% & \underline{81.75\%} \\
\bottomrule
\end{tabular}
\end{table*}

\subsection{Online A/B Testing}
\label{sec:online}
\begin{table}[h]
\centering
\caption{Online A/B testing results in production environments. Improvements are reported as percentage lifts relative to the Production Baseline. }
\label{tab:online_results}
\resizebox{\linewidth}{!}{
\begin{tabular}{llcc}
\toprule
\textbf{Scenario} & \textbf{Model} & \textbf{Like Rate} & \textbf{Total Listening Time} \\
\midrule
\multirow{5}{*}{\textbf{General}} & Baseline & -- & -- \\
& SASRec & $-0.38\%$ & $+1.00\%$ \\
& HSTU & $+0.85\%$ & $+3.05\%$ \\
& \text{\CLIMBER} (w/o SFT) & $+2.03\%$ & $+3.86\%$ \\
& \textbf{\CLIMBER} & $\bm{+4.24\%}$ & $\bm{+3.25\%}$ \\
\midrule
\multirow{2}{*}{\textbf{Genre-Specific}} & \text{\CLIMBER} (w/o SFT) & $+0.68\%$ & $-0.38\%$ \\
& \textbf{\CLIMBER} & $\bm{+4.10\%}$ & $\bm{+0.56\%}$ \\
\bottomrule
\end{tabular}
}
\end{table}

\subsubsection{Overall Performance Comparison}
We conducted a two-week online A/B test on NetEase Cloud Music, allocating 5\% of live traffic to each model variant, across two distinct production scenarios: (1) General Recommendation, which provides music recommendations retrieved from the entire item corpus; and (2) Genre-Specific Recommendation, which caters to users seeking music from a particular category. We benchmarked against the production baseline (a Transformer-based architecture) and SOTA models including SASRec~\cite{kang2018sasrec} and HSTU~\cite{zhai2024hstu}, with results summarized in Table~\ref{tab:online_results}.

In the unconstrained General scenario, \text{\CLIMBER} (w/o SFT) already outperforms strong baselines. This indicates that our pre-training objective effectively mitigates the inherent myopic prediction bias, thereby capturing long-term user interests to enhance overall satisfaction.

In the Genre-Specific scenario, the w/o SFT variant works in a simple way: the model retrieves candidates without any genre signal, and the system then keeps only those whose genre matches the target. This approach yields only marginal improvements. Analysis of traffic logs during the same A/B test period shows that it retains on average only about 40\% of the retrieved candidates matching the target genre, while CGSA raises this proportion to 77.9\%. In contrast, by feeding the target genre as the retrieval instruction, the full \text{\CLIMBER} model achieves an observed lift of $4.10\%$ in Like Rate. This demonstrates that SFT effectively equips the model with instruction-following capabilities to meet specific retrieval requirements.

Notably, the introduction of SFT further amplifies user engagement in the General scenario, achieving a Like Rate improvement of $4.24\%$—more than doubling the $2.03\%$ gain observed without SFT. In this setting, personalized instructions are constructed following the strategy in Sec.~\ref{subsubsec:pre-computed}. This empirical evidence suggests that instruction-following is not solely beneficial for Genre-Specific cases; when coupled with user-adaptive instruction construction, SFT effectively enhances the model's ability to interpret and match latent user intent, resulting in superior recommendation quality.
Regarding inference efficiency, \text{\CLIMBER} achieves an average online latency of 7.26\,ms per request, compared to 6.94\,ms for the production baseline—an overhead of less than 5\%, which is negligible relative to the substantial accuracy gains.

\subsubsection{Engagement Diversity Analysis}
\label{subsec:diversity_analysis}

\begin{table}[h]
\centering
\caption{Relative improvement in exposure diversity (Genre and Language) across user activity levels.}
\label{tab:diversity_metrics}
\begin{tabular}{lcc}
\toprule
\textbf{User Activity Level} & \textbf{Genres} & \textbf{Languages} \\
\midrule
Low        & $+0.28\%$ & $+0.09\%$ \\
% $(10, 50]$       & $-0.38\%$ & $+0.50\%$ \\
Middle      & $+1.87\%$ & $+2.28\%$ \\
% $(100, 400]$     & $+1.35$ & $+2.29$ \\
High & $+2.56\%$ & $+2.63\%$ \\
\bottomrule
\end{tabular}
\end{table}

Table~\ref{tab:diversity_metrics} presents the relative improvement in exposure diversity (Genre and Language) brought by \text{\CLIMBER}. User activity levels are categorized into three groups (Low, Middle, High) based on internal business criteria. A notable trend is that the diversity gain scales with user activity.
\text{\CLIMBER} significantly enhances the exploration breadth for high-activity users, with diversity metrics improving by over $2\%$. This demonstrates that the instruction-following capability of \text{\CLIMBER} not only boosts core business metrics, but also enables the system to better satisfy the multi-modal interests of active users.

\subsection{Case Study}
% To assess the instruction-following capability of \text{\CLIMBER}, we present a case study of a randomly sampled user with mixed interests in both Hip-Hop and Classic Hits.

% In the unconstrained setting, \text{\CLIMBER} retrieves a mixed candidate set reflecting both genres. Upon injecting explicit genre instructions like "Hip-Hop" or "Classic Hits", the model generates recommendations strictly confined to the target genre, ensuring precision. This behavior validates the model's precise instruction-following capability.

\begin{figure}[h]
  \centering
  \includegraphics[width=\linewidth]{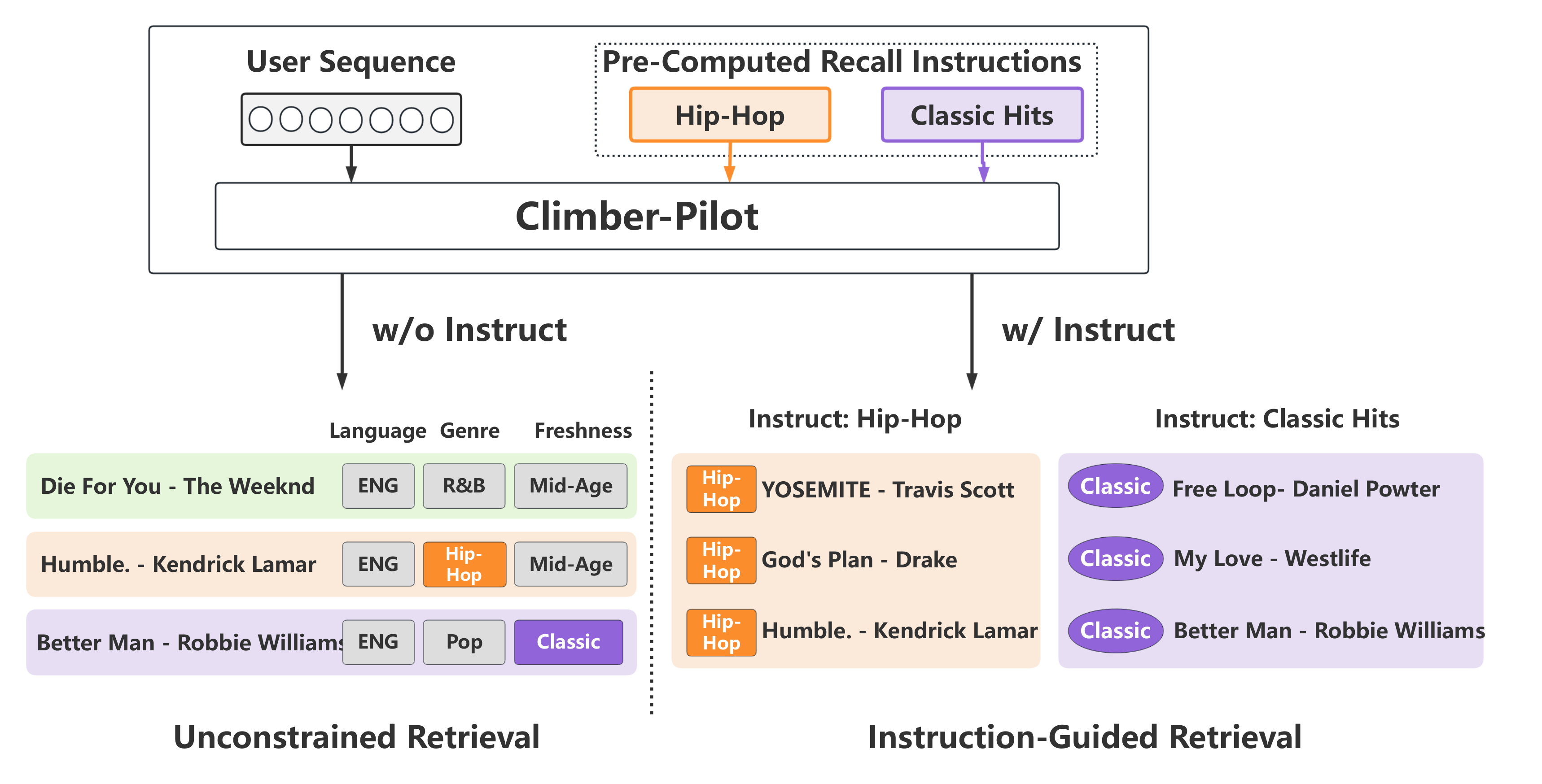}
  \caption{A case study demonstrating the instruction-following capability of \text{\CLIMBER}.}
  \Description{Case Study}
  \label{fig:case}
\end{figure}

To assess the instruction-following capability of \text{\CLIMBER}, we present a case study of a user with mixed interests in Hip-Hop and Classic Hits, as shown in Figure~\ref{fig:case}. In the unconstrained setting, \text{\CLIMBER} retrieves candidates reflecting both genres. Upon injecting genre instructions, the model generates recommendations strictly confined to the target genre. 

Crucially, this strict adherence does not compromise personalization. The instruction-guided recommendations exhibit overlap with the results from the unconstrained setting, demonstrating that \text{\CLIMBER} successfully identifies the intersection of user interests and instructional constraints.

These results validate our CGSA mechanism. By filtering out instruction-irrelevant interactions to ensure precision while selectively attending to context-relevant behaviors to maintain personalization, \text{\CLIMBER} achieves both precision and personalization within a single unified model.

\section{Conclusion}
% We propose \text{\CLIMBER}—an instruction-following generative retrieval framework that addresses both inherent myopia and controllability limitations in recommendation systems. Our two-stage training paradigm first endows the model with general retrieval capability through TAMIP, which internalizes long-horizon intent evolution without inference overhead, then equips controllable retrieval capability via CGSA. Extensive experiments demonstrate consistent improvements over state-of-the-art baselines. Online A/B testing at NetEase Cloud Music validates practical effectiveness, achieving 4.24\% improvement on Like Rate and 3.25\% improvement on Total Listening Time for general recommendation. By unifying diverse retrieval requirements into a single steerable model, \text{\CLIMBER} substantially reduces engineering complexity while establishing a paradigm for instruction-aware generative retrieval in industrial music streaming platforms.

In this work, we identified inherent myopia and instruction-following inefficiency as fundamental challenges of generative retrieval models under single-step inference and strict latency constraints in industrial recommender systems. To address these issues, we proposed \text{\CLIMBER}, a unified framework that distills long-horizon, multi-item user intent during training and incorporates retrieval instructions directly into the generative process. Extensive offline experiments and large-scale online A/B testing in a production system demonstrate that \text{\CLIMBER} consistently outperforms state-of-the-art baselines, achieving consistent improvements in core business metrics while preserving efficient single-step inference. These results highlight the practical effectiveness of \text{\CLIMBER} for deploying generative retrieval models in real-world, latency-sensitive recommendation systems.

%%
%% The acknowledgments section is defined using the "acks" environment
%% (and NOT an unnumbered section). This ensures the proper
%% identification of the section in the article metadata, and the
%% consistent spelling of the heading.

%%
%% The next two lines define the bibliography style to be used, and
%% the bibliography file.
\balance
\bibliographystyle{ACM-Reference-Format}
\bibliography{refer}

%%
%% If your work has an appendix, this is the place to put it.

\end{document}